\documentstyle[11pt,newpasp,twoside,epsf]{article}

\def\edcomment#1{\iffalse\marginpar{\raggedright\sl#1\/}\else\relax\fi}

\reversemarginpar

\begin{document}

\title{HI metal  enrichment from massive star--forming regions: sudden or delayed?}

 \author{Kunth Daniel}

\affil{Institut d'Astrophysique de Paris, 98 bis Bld. Arago, F-75014, Paris; France}

\begin{abstract}

 It now  becomes possible to perform
multiphase studies of the ISM from hot to molecular phases. H\,{\sc i}  and H\,{\sc ii}
in star--forming gas-rich galaxies seem to exhibit differences in their metal
composition in the sense that H\,{\sc i} is less enriched in nitrogen, argon and
possibly oxygen than the H\,{\sc ii} ionized gas. This observational 
evidence needs to
be confirmed on larger sample and yet remains difficult to intepret. We 
review some possible interpretations. If, on the other hand, 
metals remain for a
long time in hot gas they might ultimately  cool down and mix with the original
interstellar medium rather than being expelled into the intergalactic medium.

\end{abstract}

\section{The quest for extreme metal poor galaxies}

The epoch of galaxy formation is still an uncertain issue. The possibility
that at our present epoch some uncondensed clouds of primordial matter may 
 condense and 
form massive stars for nearly the first time remains unconclusive. 
No evidence for local  unprocessed primordial cloud of gas
has yet been given and  the discovery of 
such pristine gas would be of great importance for cosmology. Good 
places to search for  are the blue compact galaxies
(hereafter BCGs, also
named extragalactic H\,{\sc ii} regions) which are very deficient in heavy
elements as compared to our Galaxy (Kunth \& \"Ostlin 2000). 
Most BCGs are 
rich in neutral gas which might be totally or partially unprocessed 
in some cases. The true 
metallicity  of
these  unevolved systems could be even lower than that  measured 
from
their H\,{\sc ii} regions if they are self--enriched in heavy 
elements by  massive 
stars formed in the present burst. Such a self enrichment mechanism
would explain the failure to find extreme metal poor star--forming galaxies
from extensive emission-line galaxies surveys (Kunth \& Sargent 1986; 
hereafter KS86). Pantelaki \& Clayton 
(1987) dismissed this possibility from the fact that most of the ejecta 
should remain for a long time in the hot gas generated by supernov\ae\ (SN) 
events. In more general terms, Roy \& Kunth (1995) discuss mixing
processes in the interstellar medium (ISM) of gas rich galaxies and conclude 
that dwarf galaxies
are expected to show kpc scale abundance inhomogeneities.  On the other hand, 
chemodynamical models (Hensler \& Rieschick 1998), 
predict that the ISM will be well mixed and chemically homogeneous through 
cloud evaporation.

\section{Evidences for/against abundances homogeneities}

Do we measure abundance inhomogeneities?

The observational situation is still not completely clear. Few 
dwarfs have been subject to high quality studies of their chemical homogeneity.
Most dwarf irregulars seem rather homogeneous (Kobulnicky 1998) with the 
exception 
of NGC~5253 where local N/H overabundances has been attributed to localised 
pollution from  WR stars (Kobulnicky et al. 1997). 
There is also marginal 
evidence for a weak abundance gradient in the LMC  (on the scale of 
several kpc, Kobulnicky 1998).

The situation is less clear in BCGs: 
In IIZw~40, Walsh \& Roy (1993) found a factor two variation in 
the oxygen abundance while Thuan et al. (1996) report N/H local overabundances
that they attribute to stellar WR winds. However this is not true for 
all young starbursts even when WR stars are suspected or well observed
(Oey \& Shields, 2000). On the other hand
IZw~18 appears to be rather homogeneous (e.g. Skillman \& Kennicutt 1993, Vilchez \& Iglesias-P\'aramo 
1998, Legrand et al. 2000)
hence does not advocate in favor of the concept of self-enrichment (KS86). On 
the other hand recent spectroscopy of SBS0335-052 (Izotov et al. 1999) 
reveals small but significant variations in accordance (though to a much
lesser extent)  with previous results (Melnick et al. 1992). 

Arguments in favor of complete mixing lie from the observation that 
disconnected H\,{\sc ii} regions within the same galaxies have nearly the same
abundances. Six H\,{\sc ii} regions in the SMC have log O/H = 8.13 ($\pm 0.08$)
while 4 in the LMC give log~O/H = 8.37 ($\pm 0.25$) (Russell \& Dopita 1990).
Even though Dennerl et al. (2001) measure oversolar O, Ne, Mg, and Si
(all type II supernov\ae\ products) abundances in the LMC, it is not
known which fraction of the processed gas is contained in hot phase and
when this gas will condense into the H\,{\sc i} phase. The possibility that 
metallicities in the neutral gas phase are 
orders of magnitude below the H\,{\sc ii} region abundances  would be
an ultimate test of large scale inhomogeneities.

\subsection{A galaxy with extreme properties: IZw~18}

Amongst blue compact galaxies, IZw~18 has the lowest known heavy-element 
abundances in its H\,{\sc ii} regions  and is thus the best candidate for a 
young galaxy experiencing its first star formation.
The utter lack of known
galaxies with abundances smaller than  IZw~18, despite concentrated 
observational efforts  has been a puzzle. As we stressed above, 
KS86 
suggested that IZw~18 could indeed be a primordial galaxy in which the observed H\,{\sc ii} regions have been self-enriched in the current burst. 

Its youth has been questioned by Aloisi, Tosi \& Greggio (1999) using HST
optical data and in the near infrared by \"Ostlin (2000) who discovered a
well-defined population of asymptotic giant branch stars. Hovewer Hunt, Thuan
 \& Izotov (2003) argue that only 22$\%$ of the total mass could be contributed
 by older stars which in itself makes this object peculiar per se.
It also contains a relatively large amount of neutral gas (Viallefond et
al. 1987) and the main H\,{\sc ii} region of IZw~18 is associated with a 
very massive neutral cloud. The content of this large
surrounding envelope has been the subject of debates. For instance it had
been suggested that this envelope  may contain a significant reservoir of
molecular hydrogen to account for dark matter (Lequeux \& Viallefond 1980).

  In any case, as  part of this 
cloud is located on the line of sight to the ionizing star
cluster, heavy elements can be detected through their absorption lines 
against the continuum of the ionizing stars.
Spectra were first obtained with the Goddard High Resolution 
Spectrograph (\textit{GHRS}) on board the Hubble Space Telescope (HST) in the spectral
regions around the Lyman $\alpha$ line of hydrogen and the resonance line of
neutral oxygen O\,{\sc i} at 1302 \AA. Unfortunately this line has a very strong
oscillator strength ($f_{1302}=51.9\times 10^{-3}$) and becomes rapidly 
saturated, in particular  at the
large column densities 
observed ($N_{\rm H\,{\sc i}}\ga 2\times 10^{20}$cm$^{-2}$). 
Results were unconclusive and
let the problem with divergent conclusions (Kunth et al. 1994, Pettini 
\& Lipman 1995).  Van Zee et al. (1998) using a new value for the H\,{\sc i} velocity
dispersion were able to obtain an H\,{\sc i} metallicity more in the range to that in
the H\,{\sc ii} regions. Izotov, Schaerer \& Charbonnel (2001) still defend the view that UV
lines may originate from the H\,{\sc ii} regions leaving open the 
possibility that H\,{\sc i}
is pristine.

\section{Far Ultraviolet Spectroscopic measurements}

 We now enter a time  in which it  becomes possible to perform
multiphase studies of the ISM. In addition to the H\,{\sc ii} regions 
which have been
routinely studied over years from ground--based optical facilities, hot gas
studies are 
now at reach owing to X-rays satellites such as Chandra, while the 
chemical
 composition of H\,{\sc i} gas and the presence  of H$_2$ molecules are 
accessible using the HST and the \textit{FUSE}. Stellar spectroscopists
reach the possibility to  measure the atmospheric
composition of stars in local galaxies. All together the understanding of
galaxy evolution driven by the evolution of its primary constituents is now
a feasible goal.

 Given the chemically unevolved nature of some BCGs
 and the lack of organized gas dynamics
and/or spiral arms, it is unclear where and how these galaxies formed the
  molecular gas  thought to be required to produce the current generation of 
young stars. 
However, the remarkable absence of diffuse H$_2$
in the most metal-poor  starburst galaxy IZw~18 can well  
be explained by the low abundance of dust  grains, its high ultraviolet flux, 
and the low density of the H\,{\sc i} cloud surrounding the star-forming regions (Vidal-Madjar et al. 2000)

 The launch of \textit{FUSE} has provided access to the rich system 
of far-UV absorption lines that can be
seen against the  900--1200 \AA\ continuum light that is produced by ionizing
stellar clusters in BCGs.  The high spectral resolution of \textit{FUSE} allows to
analyse and disentangle the Galatic contribution giving a possibility to
retrieve the metal content of the neutral ISM of BCGs. It is now possible to
measure abundances of O, Si, Ar, Fe, P  by using several lines per ions and
applying a procedure allowing for their simultaneous fit.

\begin{figure}[h!] 
\plotfiddle{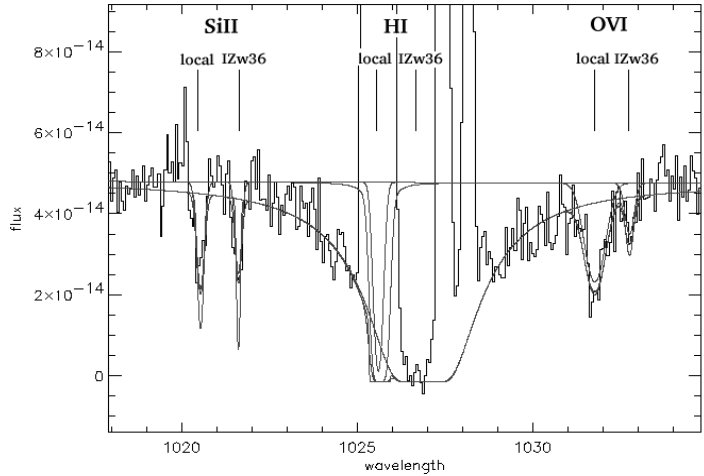}{200pt}{-90}{40}{38}{-160}{210}  
	\caption{\it{The Lyman $\beta$ line in IZw~36. The flux is in
erg~s$^{-1}$~cm$^{-2}$~\AA$^{-1}$ and the observed wavelength in \AA. The
data 
 (histogram)  are binned by a factor 3 for  display purposes. 
The thin line  represents the line profile before convolution with the
line spread function and the  thick line is the convolved profile.  The
center of the Ly$\beta$ line is blended with terrestrial  airglow emission
lines. Atomic hydrogen in the Milky Way  is invisible because of its low
column density $\sim$10$^{19}$~cm$^{-2}$. Metallic lines
from the Milky Way and IZw~36 are also visible in this spectrum.}}
 \label{fig:lyman} 
\end{figure}

H\,{\sc i} column densities are calculated using the Lyman series (Fig.~1).
So far thirty BCGs have been observed with \textit{FUSE}. The chemical composition
of four BCGs have been fully investigated: IZw~18 (Lecavelier et al. 2003
and Aloisi et al. 2003), IZw~36 (Lebouteiller et al. 2003), SBS~00335-052 (Lecavelier
et al. 2002) and Markarian~59 (Thuan et al. 2002).

Oxygen is produced by massive stars and released in the ISM through type II 
supernov\ae\ explosions and is usually taken as a reference element to
investigate the composition of the ISM in extragalactic objects. Its
derivation in ionized regions is very reliable while as we explained above
its derivation for H\,{\sc i} gas is more complicated. \textit{FUSE} allows to observe several
oxygen lines but they are either saturated, midly saturated or blended with
H\,{\sc i} Lyman lines hence in all cases difficult to interpret. For this reason
the actual results display confortable error bars and a very contrasted
situation.  For IZw~18 Lecavelier et al. find  
$\log~$(O\,{\sc i}/H\,{\sc i})=$-4.7^{+0.8}_{-0.6}$ which is  consistent with the O/H
ratio observed in the H\,{\sc ii} regions (all uncertainties are
2-$\sigma$) while Aloisi et al. (2003) report  a significantly
different value with
$\log~($O\,{\sc i}/H\,{\sc i})=$-5.4\pm$0.3  a discrepancy with the former
result that awaits for clarification (see Lecavelier et al. 2003 on this
point). 
Thuan et al. (1997) attempted to circumvent the 
problem of the O\,{\sc i} 1302~\AA\  saturation  in the case of SBS0335-052 
and reported an 
extremely metal
deficiency in the H\,{\sc i} gas with an O\,{\sc i}/H\,{\sc i} ratio
 $\la 3 \times 10^{-7}$. However in that same galaxy the O\,{\sc i}/H\,{\sc i}
ratio was subsequently found to be much higher using \textit{FUSE}
observations of O\,{\sc i} lines at shorter wavelengths and with
smaller oscillator strengths (Lecavelier  et
al. 2002). In
fact, the hypothesis of an O\,{\sc i} line at 1302~\AA\
unsaturated was simply erroneous. It is noteworthy that O\,{\sc i}/H\,{\sc i} ratios in
SBS0035--052 and Mrk~59 are similar to that of IZw~18. This may suggest that
the metallicity of the surrounding H\,{\sc i} cloud is unrelated to the metallicity
of the H\,{\sc ii} regions. Coincidental or not this remains to be observationally
supported using a larger sample. In any case Mrk 59 is the only object for
which the difference between the H\,{\sc i} and the H\,{\sc ii} regions is significant.

A more secure issue would come from the measurement of unsaturated lines
such as the S\,{\sc ii} ${\lambda 1256}$ multiplet, unfortunately out of the \textit{FUSE}
spectral window. Can we derive total abundances using other elements?
It is of course necessary to pay attention to the ionization states of each
species under study.  Nitrogen (Fig.~2) is of twofold interest:
\begin{itemize}
\item  its presence in the neutral regions is well established: the ionization potential of N\,{\sc i}
is 14.53 eV and it can be shown that the applied correction to account for
the possible presence of ionized N in the neutral gas is only 0.05 dex (Sofia
\& Jenkins, 1998),
\item many of the N\,{\sc i} lines are
unsaturated hence weakly depend on the turbulent velocity of the gas. 
\end{itemize}\par

\begin{figure}[h!] 
\plotfiddle{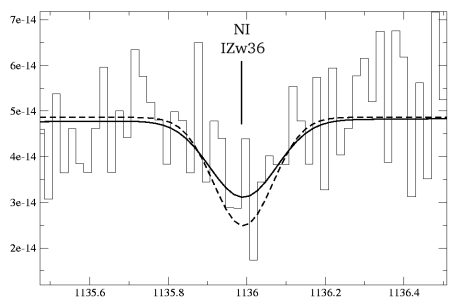}{130pt}{0}{25}{24}{-105}{0}  
      \caption{\it{Fit of the $\lambda$1134.98 line in IZw~36.  
See Fig.~1 for a description of the plot. Data are binned by a factor 2 
for  display purposes. 
The solid line is for log~(N\,{\sc i}/H\,{\sc i})=$-$6.88 and shows the best
      fit. 
The dashed line is for log~(N\,{\sc i}/H\,{\sc i})=$-$5.63 (assuming the same 
N/H ratio that in the ionized gas) and does not fit well the data at 
3-$\sigma$.}} 
         \label{fig:nitrogen} 
\end{figure}

So far the extended neutral regions of gas rich galaxies seem to exhibit
abundances differences between their neutral and their ionized has. If 
unfortunately oxygen gives unconclusible results the most unambiguous  but  
puzzling evidence for such a chemical difference comes from nitrogen and
argon (Lebouteiller, this conference and Lebouteiller et al. 2003). The interpretation remains difficult.
We  can   identify several effects  responsible for these deficiencies:\\ 
$-$ \textbf{The presence of an unprocessed neutral gas} in the line of sight  
could lower all the abundances in the neutral region.\\ 
$-$ \textbf{Depletion on dust grains} can account for a possible iron deficiency but  is not likely to affect neither nitrogen or argon.\\ 
$-$ \textbf{Prompt metal enrichment} (self pollution) can be responsible for 
nitrogen and argon  overabundance in the ionized gas.  \\

Self-enrichment by SNe could account for the excess nitrogen observed in H\,{\sc ii} regions w.r.t. that of the H\,{\sc i} gas. On the other hand the low abundance pattern
of the interstellar medium suggests ancient star-formation activity with an
age of around a Gyr or a continuous low level star formation activity over
several Gyr (Legrand, 2000; Aloisi et al. 2001). Such abundances differences
between the two gas phases are uneasy to explain by Tenorio-Tagle's model
(1996)  in which hot gas cools and later on settles down onto the disk. At 
present, the overall picture remains unclear and it is possible
that a larger sample will help to determine more accurately the  oxygen
composition  of the H\,{\sc i}. A way out might well be to use instead 
phosphorus asymmetry a tracer of oxygen abundance, easier to measure 
in \textit{FUSE} spectra
(Lebouteiller, 2003).

\section{Metals from star-forming dwarfs: retention or ejection?}

As they end their lives  massive stars  explode as a 
supernov\ae. The energy output from a SN is over a short period,
comparable to that of a whole galaxy.
In a galaxy with a high local star formation rate, the collective action of 
supernov\ae\ may lead to a galactic superwind, which may cause loss of gas. 
Stellar winds can also contribute to the energetics of the ISM
at the very early stage of a starburst (Leitherer et al. 1992). The relative
importance of stellar winds  compared to SNe increases with metallicity. 
A continuous wind  proportional to the star formation rate  has
been applied in models predicting the evolution of starburst galaxies. 
But since different elements are produced on different timescales, it has
been  proposed that only certain elements are lost (or in different
proportions) hence reducing the effective net yield of those metals as 
compared to a simple chemical evolution model (Matteucci \& Chiosi 1983, 
Edmunds 1990). 
The SNe involved in such a wind are likely to be of type II because type Ia 
SNe explode in isolation and will less likely trigger chimneys from which
metals can be ejected out of the plane of a galaxy. In this framework O and
 part of Fe  are lost while He and N 
(largely produced by intermediate stars) are not. This would result in a 
cosmic dispersion in element ratios such as N/O between galaxies that 
have experienced mass loss and those that have not.
In a dwarf galaxy which has a weaker gravitational potential, these effects 
may result in gas loss from the galaxy unless as we argue below the presence
of a low HI density halo acts as a barrier. 

Galactic winds have been observationally investigated in dwarf 
galaxies (e.g. Marlowe et al. 1995; Martin 1996, 1998) and more recently with
the advent of the Chandra satellite. Lequeux et al. (1995), Kunth et al.
(1998) and Mas-Hesse et al. (2003) have shown that the escape of 
the Ly$\alpha$ photons  
in star-forming galaxies strongly depends on the dynamical properties of  
their interstellar medium. The Lyman alpha profile in the BCG Haro~2 indicates a superwind of at least 
200 km/s, carrying a mass of $\sim 10^7 M_{\odot}$, which can
be independently traced from the H$\alpha$ component (Legrand et al. 1997).
However,  high speed winds do not necessary carry a lot of mass.
Martin (1996)  argues that a bubble seen in IZw~18 (see also Petrosian et al.
1997) will ultimately blow-out together with its hot gas component. 
Although little is known  about the interactions between the evolving 
supernova remnants, massive stellar bubbles and the ISM it is possible that 
an outflow takes the fresh metals with it and in some cases leaves a galaxy 
totally cleaned of gas. This scenario clearly contradicts the self enrichment
picture - unless a fraction of hot gas rapidly cools down -. But since we see
some hot gas outside of the H\,{\sc ii} regions the
question remains whether  this gas will leave a galaxy or simply stay 
around in the halo? 

 Model  calculations  developed by Silich \& Tenorio-Tagle (1998)
 predict that  superbubbles in   amorphous  dwarf galaxies must have already 
undergone blowout and are presently
evolving into an extended low-density halo. This should inhibit the loss 
of the  swept-up and processed matter into the IGM. Recent
 Chandra X-rays  observations of young starbursts 
 indicate some  possible metal losses 
 from  disks (Martin, Kobulnicky \& Heckman 2002) but not in  the 
case of NGC~4449 with an extended H\,{\sc i} halo of around 40~kpc 
(Summers et al. 2003).
In a  starburst galaxy, newly processed elements are produced within a 
region of $\sim 100$ pc size. During the supernova phase the continuous energy 
input rate from coeval starbursts or continuous star--forming episodes, 
maintains the temperature of the ejected matter  
above the recombination limit ($T \sim 10^6$ K) allowing superbubbles to 
reach dimensions in excess of 1 kpc. 

  The mechanical energy released during a starburst episode  accelerates 
the interstellar medium gas and generates gas flows. The 
properties and evolution of these flows ultimately determine the fate of the 
newly formed metals and the manner they  mix with the original interstellar 
medium.  
 The presence of outflows may indicate, at first, that supernova  products 
and even the whole of the  
interstellar medium may be easily ejected from the host dwarf systems,  
causing the contamination of the intra-cluster medium (Dekel \& Silk 1986; 
De Young \& Heckman 1994). This type of assumption is currently blindly used by
cosmologists in their model calculations.
However the indisputable presence of metals in galaxies implies that  
supernova products are not completely lost in all 
cases (Silich \& Tenorio-Tagle 2001). Legrand et al (2001) have compared 
Silich \& Tenorio-Tagle (2001) theoretical estimates with some 
well--studied starburst galaxies.  
They have  worked out three different possible star 
formation history scenarios that assume either a very young coeval 
starburst or  extended phases of 
star formation of 40 Myr or 14 Gyr and inferred  the expected  energy 
input rate  for  each galaxy using the  H$\alpha$ luminosity  and/or 
the observed metallicity. 
Values of the derived mechanical energy injection
rate in the three considered cases were compared with the  hydrodynamical  
models predictions of  Mac Low \& Ferrara (1999) and Silich \& Tenorio-Tagle 
(2001). Detailed calculations and accompagnying figures are given in 
Legrand et al. (2001); we only present in Fig.~3 the resulting plot for the 
extended phases of star formation of 40 Myr that might be the most realistic 
case for starburst galaxies.

\begin{figure}[h!] 
\plotfiddle{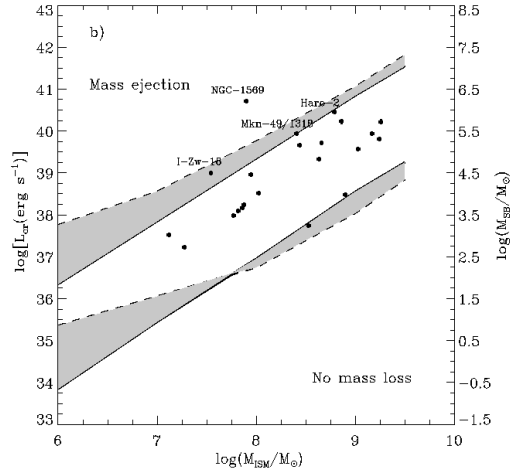}{200pt}{0}{30}{30}{-105}{0} 
 
\caption{\it{ Log of the critical mechanical luminosity  
(left-side axis) and Mass of the star cluster $M_{SB}$ (right-side axis),
required to eject matter from galaxies as a function of $M_{ISM}$. Lower
limit estimates are shown for galaxies with extreme ISM density
distributions: flattened disks (lower two lines), and spherical galaxies
without rotation (upper lines), for two values of the intergalactic
pressure $P_{IGM}/k$ = 1 cm$^{-3}$ K (solid lines) and $P_{IGM}/k$ = 100
cm$^{-3}$ K (dashed lines).  Plot refers to Fig1-b in Legrand et al. (2001). }}
\end{figure}

The net result (see Fig.~3) is that  all galaxies  lie  
above the lower limit first derived by  Mac Low \& Ferrara (1999)
for the ejection of metals out of flattened  
disk-like ISM density distributions energized while most are BELOW the 
limit for the low density  halo picture. 
Thus the mass of the extended low density halo  efficiently acts as the 
barrier against metal ejection into the IGM.

 Disk-like models clearly require less energy 
to eject their metals into the intergalactic medium because the amount of 
blown out interstellar gas that can open a channel into the
intergalactic medium is much smaller than in the spherically symmetric limit, 
where  all of the metal-enriched ISM has to be accelerated to reach the
galaxy boundary.
Predicted haloes, despite acting  
as a barrier to the loss of the new metals, have rather low densities  
($<n_{halo}> \sim 10^{-3}$ cm$^{-3}$) and thus have a long recombination  
time ($t_{rec} = 1/(\alpha~n_{halo}$); where $\alpha$ is the recombination  
coefficient) that can easily exceed the lifetime of the  H\,{\sc ii} region  
($t_{HII}$ = 10$^7$ yr) developed by the starburst.  In such a case,  
these haloes may remain undetected at radio and optical frequencies 
 until large volumes are collected into the  
expanding supershells.\par

\section{Conclusions}

\begin{itemize}
\item On average  H\,{\sc ii} gas across a given galaxy has a fairly uniform
composition. Some local inhomogeneities might be due to local stellar
enrichment although this picture is unsettled yet.
\item Several heavy elements are found to be less abundant in the surrounding
 H\,{\sc i} gas of starburst galaxies than that measured
from their  H\,{\sc ii} regions. This observational evidence is rather well
established for nitrogen and argon. Oxygen turns out to be difficult to
quantify but using  P\,{\sc ii} lines and P/O one is led to conclude that oxygen  in
 H\,{\sc i} might also be less abundant than in the  H\,{\sc ii} region
\item The above chemical differences observed between   H\,{\sc i} and  H\,{\sc ii} gas
phases, if real are uneasy to interpret. They may be due to the presence of
unprocessed amount of gas in the line of sight, to self enrichement effects
within the  H\,{\sc ii} regions or else. A further statistical study involving larger
sample of BCGs and Giant  H\,{\sc ii} regions is under way (Lebouteiller's
thesis)
.
\item The presence of metals in BCGs that are gas-rich and star-bursting
objects points towards the presence of extended HI haloes.
\end{itemize}

\end{document}